\documentclass[aps,
pra,
superscriptaddress,
reprint
]{revtex4-2}
\usepackage{quantikz}
\usepackage{graphicx}
\usepackage{dcolumn}
\usepackage{bm}
\usepackage{lipsum}
\usepackage{siunitx}
\usepackage{physics}
\usepackage{amssymb}
\usepackage{placeins}
\usepackage{comment}
\usepackage{tcolorbox} 
\usepackage{braket} 
\usepackage{xcolor} 
\usepackage{amsmath} 
\usepackage{bbm} 
\usepackage[caption=false]{subfig} 
\definecolor{darkgreen}{RGB}{0,150,0}
\usepackage[normalem]{ulem} 
\usepackage[colorlinks,linkcolor=blue,urlcolor=blue, citecolor=blue]{hyperref}
\begin{document}

\title{Photonic Violation of Wigner's Inequality}

\author{Maximilian Rottensteiner}
\email{maximilian.rottensteiner@student.tuwien.ac.at}
\affiliation{Institute for Quantum Optics and Quantum Information - Vienna, Austrian Academy of Sciences, Boltzmanngasse 3, 1090 Vienna, Austria}

\author{Dorian Schiffer}
\affiliation{Institute for Quantum Optics and Quantum Information - Vienna, Austrian Academy of Sciences, Boltzmanngasse 3, 1090 Vienna, Austria}
\affiliation{Atominstitut, Technische Universität Wien, Stadionallee 2, 1020 Vienna, Austria}

\author{Tobias Pausch}
\affiliation{Institute for Quantum Optics and Quantum Information - Vienna, Austrian Academy of Sciences, Boltzmanngasse 3, 1090 Vienna, Austria}

\author{Alois Mair}
\affiliation{Institute for Quantum Optics and Quantum Information - Vienna, Austrian Academy of Sciences, Boltzmanngasse 3, 1090 Vienna, Austria}

\author{Anton Zeilinger}
\email{anton.zeilinger@oeaw.ac.at}
\affiliation{Institute for Quantum Optics and Quantum Information - Vienna, Austrian Academy of Sciences, Boltzmanngasse 3, 1090 Vienna, Austria}

\begin{abstract}
    
    Teaching quantum mechanics is challenging, not least because the theory often conflicts with our classical worldview. Quantum correlations in particular are notoriously counter-intuitive. Their non-classical behavior is typically revealed through Bell-type inequalities.
    Among these, Wigner’s Inequality constitutes a particularly accessible test, as it relies on minimal set-theoretic assumptions. In this pedagogical paper, we derive  Wigner’s Inequality, describe a quantum-optical setup to experimentally violate it, and provide access to the raw data, enabling students and instructors to perform their own analyses. Our measured data shows clear violations of Wigner’s Inequality, directly illustrating the non-classical nature of quantum correlations. By connecting theory, experiment, and data analysis, this paper equips educators with a resource for engaging students in authentic scientific practice and developing a deeper understanding of quantum systems.
\end{abstract}


 \maketitle
 
\section{Introduction}   
    Since the early days of quantum mechanics,
    researchers like Albert Einstein, Boris Podolsky, Nathan Rosen \cite{EPR}, John S. Bell \cite{Bell}, Niels Bohr \cite{Bohr1935} and Eugene Wigner \cite{Wigner} have been puzzled by the stark deviation of quantum physics from classical intuitions. Yet they did not merely question phenomena such as entanglement; they also translated these conceptual tensions into precise mathematical form by deriving no-go theorems that distinguish quantum behavior from classical predictions.

    While one generation of quantum physicists used these theorems to theoretically describe and experimentally test seemingly counterintuitive quantum effects \cite{Aspect1981,Aspect1982a,Aspect1982b, Freedman1972,Greenberger1990,Svetlichny1990,Leggett2003,Giustina2015}, these days said theorems play a central role in emerging quantum technologies \cite{Wiesner1983, Bennett_Brassard1984, Ekert1991, liao2018, Pirandola2020, Usenko2025}. They are also widely used to introduce the next generation of physicists to quantum correlations. Most prominently, Bell's theorem \cite{Bell}, often in the guise of the CHSH inequality \cite{CHSH}, is a standard topic in courses on quantum physics and quantum information. Understanding where classical models fail is a key step toward understanding quantum mechanics. Bell's theorem, however, is not the only route to that insight.

    Wigner’s Inequality (WIE) provides an alternative framework for studying the counterintuitive behavior of quantum correlations, without requiring the full apparatus of Bell’s theorem. Formally, WIE follows from a local hidden-variables (LHV) model derived using a set-theoretic argument with minimal assumptions, making it particularly well suited for introductory  physics courses. Moreover, its predictions are easily tested using conventional photonic techniques.
    
    In this work, we present an accessible derivation of WIE and its experimental test using polarization-entangled photons generated in a down-conversion-based source. Our measurements show clear deviations from the predictions of the LHV model, highlighting the non-classical nature of the observed correlations. Furthermore, they demonstrate the angular dependencies and the symmetry properties of the correlations. We describe the experimental setup and the data analysis in detail.
    
    This paper enables students to engage with the complete scientific process, from hypothesis and modeling, to experiment and data analysis, culminating in the potential falsification of the classical model in the sense of Karl Popper \cite{popper2002logic}. As such, it offers an effective and accessible pathway for developing both an intuition for quantum mechanics and a rigorous scientific mindset.

\section{Motivation}

Imagine two distant observers, potentially spacelike separated from each other. Each observer repeatedly receives particles on which they perform measurements using devices with different settings, recording the corresponding outcomes. After completing their experiments, they meet and compare their notes. This is when they discover that their outcomes are correlated: apparently, the result obtained by one observer is tied to the outcome of the measurement performed by the other.

How can these correlations be explained? Classical intuition suggests that the measurement outcomes were already determined prior to detection and stored in some hidden degree of freedom of the particles. Anything else seems to conflict with our notions of \textit{realism} and \textit{locality}.

    \begin{tcolorbox}[colback=gray!5, colframe=black!60, title=Realism and Locality]  
        According to \textit{realism} all outcomes are defined prior to their measurement \cite{PhysRevLett.99.210406}. \textit{Locality} is the principle that physical influences are confined to nearby regions of space-time and cannot propagate faster than the speed of light \cite{Lambare2022}. 
    \end{tcolorbox}
    
\pagebreak
However, can such \textit{local hidden-variables} models explain the observed correlations? Given a specific LHV model -- and we shall present one in the next section -- this question can be decided experimentally. In general, this is done by deriving bounds on the strength of the correlations that the LHV model would admit. If we find correlations in the lab that violate these boundaries, then at least one of the assumptions used to derive the model must be rejected. This may force us to depart from a classical worldview and embrace \textit{non-locality}.
    
      \begin{tcolorbox}[colback=gray!5, colframe=black!60, title=Local Hidden Variables]  
            Conceptually, LHVs posit pre-existing, unseen instructions, which dictate the individual behavior of the particles under consideration. In principle, one may conceive hidden variables that predetermine not only the measurement outcomes but even the measurement settings, although the latter are usually assumed to be chosen independently. Within this picture, all observed correlations are therefore fixed by the hidden parameters in advance.  
        \end{tcolorbox}
        
\section{Wigner’s Inequality}
\label{sec:Wigner's Inequality}
    Following Wigner's original derivation \cite{Wigner}, we consider a two-particle system. Each particle is distributed to an observer, commonly dubbed Alice and Bob. Each observer may perform projective measurements on their particle using one out of three possible settings. For the first particle these settings are $\hat{a}$, $\hat{b}$, $\hat{c}$, and likewise $\hat{a'}$, $\hat{b'}$, $\hat{c'}$ for the second particle. Each measurement can return either a $+$ (plus) or a $-$ (minus) outcome. 
    
    Wigner’s approach to recovering local realism is to model the measurement outcomes as being drawn from sets of predefined values. Each particle carries predetermined outcomes $+$ or $-$ for all possible measurement settings $\hat{a}$, $\hat{b}$ and $\hat{c}$ encoded in hidden variables.
    
    For testing this LHV model, we assume each individual particle has predefined outcomes $+$ or $-$ for each of the possible measurement settings. \textit{Assuming perfect anti-correlated outcomes}, this gives eight sets, which are combinations of measurement settings. Each set is a row in Table \ref{tab:Possible sets of measurements and their probabilities} and has a probability $p_i$ to occur. Particles from the $i$th set with $i \in [1,8]$, and the population $N_i$ have a probability of $p_i = N_i/\sum_{j = 1}^8 N_j $ to appear in the experiment.  

        \begin{table}[ht]
            \centering            
            \begin{tabular}{|c|ccc|ccc|}
            \hline
            \multicolumn{1}{|c}{Probability} & \multicolumn{3}{|c|}{Alice} & \multicolumn{3}{c|}{Bob} \\ 
                {per set}     & $\hat{a}$       & $\hat{b}$      & $\hat{c}$      & $\hat{a'}$      & $\hat{b'}$      & $\hat{c'}$     \\
            \hline
            $p_{1}$                   & $+$       & $+$      & $+$      & $-$      & $-$      & $-$     \\
            $p_{2}$                   & $+$       & $+$      & $-$      & $-$      & $-$      & $+$     \\
            $p_{3}$                   & $+$       & $-$      & $+$      & $-$      & $+$      & $-$     \\
            $p_{4}$                   & $+$       & $-$      & $-$      & $-$      & $+$      & $+$     \\
            $p_{5}$                   & $-$       & $+$      & $+$      & $+$      & $-$      & $-$     \\
            $p_{6}$                   & $-$       & $+$      & $-$      & $+$      & $-$      & $+$     \\
            $p_{7}$                   & $-$       & $-$      & $+$      & $+$      & $+$      & $-$     \\
            $p_{8}$                   & $-$       & $-$      & $-$      & $+$      & $+$      & $+$     \\
            \hline
            \end{tabular}
            \caption{List of all possible sets: A set consists of the measurement results $+$ or $-$ for all measurement settings $\hat{a}, \hat{b}, \hat{c}$ that Alice and Bob individually can choose from. Each row contains one set and its probability $p_i$. Assuming anti-correlation $2^3=8$ logical combinations are possible.}
            \label{tab:Possible sets of measurements and their probabilities}
        \end{table}

    Alice and Bob each conduct one measurement at a time. For instance, Alice may measure $\hat{a}$, while Bob chooses $\hat{b'}$. Assuming, for this example, an outcome of $+$ for both measurements, there are two possible scenarios: Either the ($+$ $-$ $+$; $-$ $+$ $-$) = $p_{3}$ or the ($+$ $-$ $-$; $-$ $+$ $+$) = $p_{4}$ set could produce this measurement outcome. The joint probability for either set to occur is
        \begin{equation}
            \label{joint probability of p^ab'}
            P_{++}^{\hat{a}\hat{b'}} =  p_{3} + p_{4}.    
        \end{equation}
    This can be repeated for the measurements of $\hat{b}$ and $\hat{c'}$ and for $\hat{a}$ and $\hat{c'}$, giving similar joint probabilities:
        \begin{equation}
            \label{joint probability of p^bc'}
            P_{++}^{\hat{b}\hat{c'}} =  p_{2} + p_{6}, 
        \end{equation}
        \begin{equation}
            \label{joint probability of p^ac'}
            P_{++}^{\hat{a}\hat{c'}} =  p_{2} + p_{4}.
        \end{equation}
    Using these three joint probabilities we construct \textit{Wigner's Inequality} WIE:
        \begin{equation}
            \label{Wigner_inequality}
            P_{++}^{\hat{a}\hat{b'}} + P_{++}^{\hat{b}\hat{c'}} \geq P_{++}^{\hat{a}\hat{c'}}.
        \end{equation}
    For practicality we can denote a ''Wigner value'',
        \begin{equation}
            \label{Wigner_value}
            \mathcal{W}=P_{++}^{\hat{a}\hat{b'}} + P_{++}^{\hat{b}\hat{c'}} - P_{++}^{\hat{a}\hat{c'}},
        \end{equation}
    and conclude from the relations above that $\mathcal{W}=p_3+p_6$. As all probabilities are non-negative, this leads to Wigner's Inequality reducing to
        \begin{equation}
            \label{Wigner_inequality_reduced}
            \mathcal{W} \geq 0
        \end{equation}
  The restriction to positive values of $\mathcal{W}$ provides a straightforward bound any classical LHV theory of the above type must satisfy.
  
    Notice, however, that deriving bounds for a given LHV model typically involves several tacit assumptions that can introduce \textit{loopholes}, preventing an unambiguous conclusion that quantum correlations have been observed even when an inequality is violated \cite{Giustina2015}. For example, Bell-like inequalities are subject to the locality loophole, which dictates spacelike separation between Alice’s and Bob’s measurement stations. In the case of WIE, the assumption of perfect anti-correlation used in its derivation constitutes an additional loophole. Because perfect anti-correlation cannot be achieved experimentally, practical implementations generally employ modified inequalities or supplementary arguments to account for this limitation \cite{Castelletto2003, Bovino2003, Bovino2008, Plick2015}. For practical reasons, in our demonstration of a violation of WIE, we do not address these loopholes, but we recommend discussing their implications as well as possible strategies for their closure with students.
  
  Similarly to the form of Wigner's Inequality in \eqref{Wigner_value}, one can derive analogous ones for other pairs of measurements, such as those corresponding to the $-$ (minus) outcome, or for permutations of the measurement settings.
  
  \begin{tcolorbox}[colback=gray!5, colframe=black!60, title=Exercise 1]
  These inequalities behave identically, and interested readers are encouraged to verify this.
  \end{tcolorbox}
    
\subsection{Quantum mechanical probabilities for a photonic system} 
    Now, we re-calculate the lower bound of $\mathcal{W}$ under the assumption that the probabilities $p_i$ are described by quantum mechanics instead of LHVs. 
  
\begin{figure}[htbp]
            \centering

            \begin{minipage}{0.48\columnwidth}
                \centering
                \includegraphics[width=\linewidth]{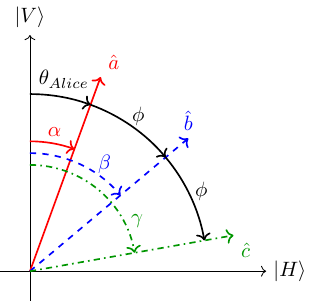}
                a) Measurement basis of Alice
                \label{fig:figure1}
            \end{minipage}
            \hfill
            \begin{minipage}{0.48\columnwidth}
                \centering
                \includegraphics[width=\linewidth]{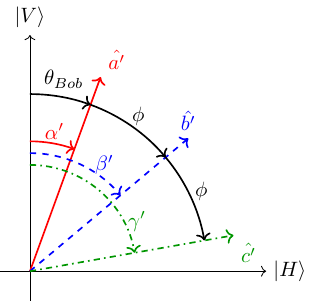}
                b) Measurement basis of Bob
                \label{fig:figure2}
            \end{minipage}
            
                      
            \caption{Visualization of the measurement settings: $\ket{H}$ and $\ket{V}$ denote horizontal and vertical polarization, respectively. The unit vectors \textcolor{red}{$\hat{a}$}, \textcolor{blue}{$\hat{b}$}, and \textcolor{darkgreen}{$\hat{c}$} represent distinct measurement directions with corresponding angles \textcolor{red}{$\alpha$}, \textcolor{blue}{$\beta$}, and \textcolor{darkgreen}{$\gamma$}. $\theta_{\mathrm{Alice}}$ and $\theta_{\mathrm{Bob}}$ denote the absolute measurement angles, while $\phi$ denotes the relative angle between the measurement directions.}
            \label{fig:Visualisation of the measurement basis}
        \end{figure}

    We utilize polarization entangled photons as our quantum system, which exhibits the required anti-correlation, while the three measurement settings correspond to arbitrary polarization filter angles placed in front of the detectors. The $+$ and $-$ measurement outcomes correspond to transmission and rejection of photons at the polarizers. The set of polarization projectors suited to test WIE is depicted in Figure \ref{fig:Visualisation of the measurement basis}.

    Using the Dirac Notation, a measurement in direction $\hat{a}$, as seen in Figure \ref{fig:Visualisation of the measurement basis}, can be expressed as a projection onto state
        \begin{equation}
            \ket{\hat{a}} = \sin{\alpha} \ket{H} + \cos{\alpha} \ket{V},
            \label{measurement directions as Kets}
        \end{equation}
     where $\ket{H}$ and $\ket{V}$ denote horizontally and vertically polarized light. 
     \begin{tcolorbox}[colback=gray!5, colframe=black!60, title=Exercise 2]
     One may construct all remaining projectors, $\ket{\hat{b}} \bra{\hat{b}}$ and $\ket{\hat{c}} \bra{\hat{c}}$, accordingly.
     \end{tcolorbox}
     Joint measurements are projections on their tensor product
        \begin{equation}
            \label{combined states ab'}
                \ket{\hat{a}\hat{b'}} = \ket{a} \otimes \ket{b'}.
        \end{equation}
  
     While deriving WIE, we assumed perfect anti-correlation whenever both observers choose the same measurement settings. Hence, we need to produce the photon pairs in the singlet Bell state \eqref{eq:singlet}, which describes perfect anti-correlation for all choices of the absolute angle $\Theta$ \cite{NielsenChuang2010}.
         
    Finally, the quantum mechanical probability can be calculated using Born's Rule \cite{Born1926}, where the measurements are projected onto the source state and squared: 
         \begin{equation}
            \label{projection}
            \begin{aligned}
                P_{++}^{\hat{a}\hat{b'}}=& \lvert\braket{\hat{a}\hat{b'}{|\psi^-}}\rvert^2 = \lvert\bra{\hat{a}\hat{b'}}\frac{1}{\sqrt{2}}(\ket{HV}-\ket{VH})\rvert^2 \\
                =& \lvert\frac{1}{\sqrt{2}}(\sin{\alpha}\cos{\beta'} - \cos{\alpha}\sin{\beta'})\rvert^2 \\
                =& \lvert\frac{1}{\sqrt{2}}(\sin{(\alpha-\beta')})\rvert^2 = \frac{1}{2}\sin^2{\phi_{\alpha\beta'}},
            \end{aligned}       
        \end{equation}
    where $\phi_{\alpha\beta'}$ is the relative angle between the measurement directions $\hat{a}$ and $\hat{b'}$. Computing the other probabilities similarly, WIE becomes
        
        \begin{equation}
            \label{eq:Wigner_inequality_equ_sin}
             \mathcal{W} = \frac{1}{2} \left( \sin^2{\phi_{\alpha\beta'}} + \sin^2{\phi_{\beta\gamma'}} - \sin^2{\phi_{\alpha\gamma'}} \right) \geq 0.
        \end{equation}
        
    This $\mathcal{W}$ is the quantum mechanical prediction of the \emph{Wigner Value} \eqref{Wigner_value}.
    From this $\mathcal{W}$, it is easy to identify the measurement settings to see a maximum deviation between  LHV and quantum predictions, to experimentally compare the competing theories. Any observation of negative Wigner values constitute a violation of the Wigner Inequality WIE,  
        \begin{equation}
            \label{eq:Wigner_bound_quant}
            \mathcal{W}
            \not \ge 0,
        \end{equation}
    ruling out classical predefined correlations, in agreement with quantum mechanical predictions.

\subsection{Predicted violations of Wigner’s Inequality -- an extremal analysis.} 
    \label{sec:Expected Violations of Wigner's inequality}
    The Wigner value $\mathcal{W}$ depends on the geometry of Alice and Bob’s measurement bases, essentially the absolute angle $\theta$ and the relative angle $\phi$ (\textit{see Figure \ref{fig:Visualisation of the measurement basis}}). 
    The measurement settings are defined by polarization filter angles restricted to a $90^\circ$ interval. For three settings the strongest violations arise with a symmetric relative spacing of $\phi = \frac{\pi}{6} = 30^\circ$.
    This selection is invariant under absolute angle rotations, as these correspond to global unitary transformations of the system \cite{griffiths2018introduction}.
    
    Two special cases arise from this.
    First, the symmetric case, where the absolute angles of Alice's and Bob's measurement bases are identical, $\theta_{Alice} = \theta_{Bob}$, using Equation \ref{eq:Wigner_inequality_equ_sin}, the model predicts an extrema at relative angles of
        \begin{equation}
            \label{Symmetric_case_angles}
            \phi_{\alpha\beta'} = \phi_{\beta\gamma'} = \frac{1}{2}\phi_{\alpha\gamma'} = \frac{\pi}{6} = 30^\circ,
        \end{equation}
    with a predicted maximal negative violation of $\mathcal{W} = -\frac{1}{8} = - 0.125$.
 
    Second, the asymmetric case, where the absolute angles of Alice and Bob differ, $\theta_{Alice} \neq \theta_{Bob}$, we evaluate Equation \ref{eq:Wigner_inequality_equ_sin} over a range of angles, and find that the strongest violation occurs for a $15^\circ$ difference between the absolute angles $\theta_{Alice}$ and $\theta_{Bob}$,
        \begin{equation}
            \label{Asymmetric_case_angles}   
            \phi_{\alpha\beta'} = \phi_{\beta\gamma'} =  \frac{\pi}{12} = 15^\circ ; \;
            \phi_{\alpha\gamma'} =  \frac{\pi}{4} = 45^\circ  .
        \end{equation}
    This leads to a theoretical maximal negative violation $\mathcal{W} = \frac{1}{4} - \frac{\sqrt{3}}{4} \approx -0.183$ for the asymmetric case. 

    \begin{tcolorbox}[colback=gray!5, colframe=black!60, title=Bell States]
    \setlength{\abovedisplayskip}{4pt} 
    \setlength{\belowdisplayskip}{4pt}
    \setlength{\abovedisplayshortskip}{4pt}
    \setlength{\belowdisplayshortskip}{4pt} 
    One can define a basis for the two-qubit Hilbert space composed of maximally entangled states, the Bell states. For historical reasons, we distinguish the singlet state,
        \begin{equation}
             \ket{\psi^-} = \frac{1}{\sqrt{2}}(\ket{HV}-\ket{VH}),
             \label{eq:singlet}
         \end{equation}
    from the three triplet states
        \begin{equation}
            \ket{\psi^+} = \frac{1}{\sqrt{2}}(\ket{HV}+\ket{VH}),
        \end{equation}

        \begin{equation}
            \ket{\phi^-} = \frac{1}{\sqrt{2}}(\ket{HH}-\ket{VV}),
        \end{equation}
        
        \begin{equation}
            \ket{\phi^+} = \frac{1}{\sqrt{2}}(\ket{HH}+\ket{VV}).
        \end{equation}
    While the singlet exhibits perfect anti-correlations in any common polarization basis of Alice and Bob, the triplet states show basis-dependent correlations that vary with the chosen measurement basis.
    \end{tcolorbox}
\section{Experimental Setup}

        \begin{figure*}[t]  
            \centering
            \includegraphics[width=\textwidth]{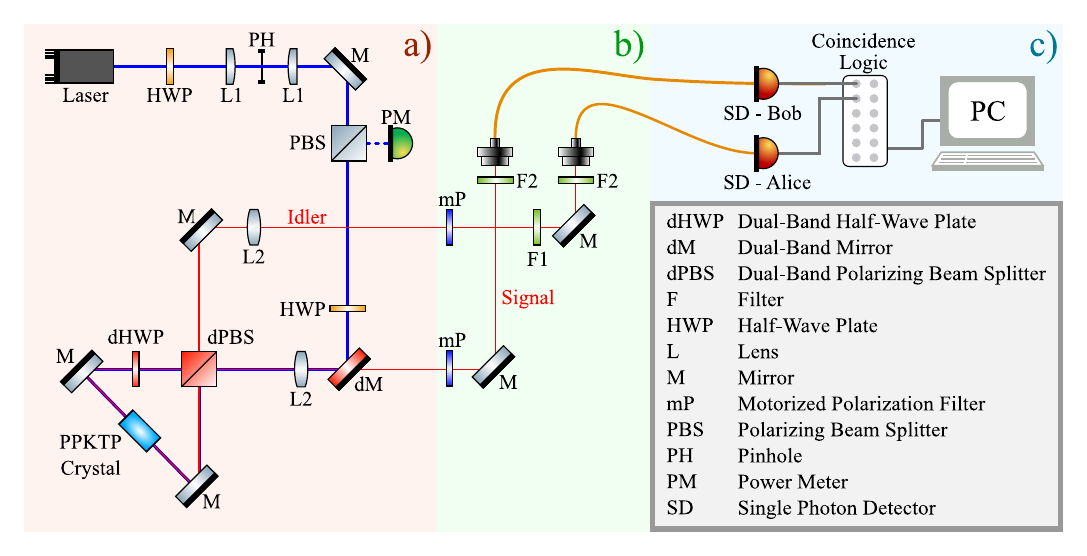}
            \caption{Schematic of the optical setup: A laser with \SI{405}{nm} wavelength is utilized to generate photon pairs at \SI{810}{nm} in a Sagnac-type photon source. These photons are measured each with a motorized polarization filter F2 and a single photon detector SD. The resulting coincidences between Alice and Bob are counted by a coincidence logic. Figure created with \cite{ComponentLibrary}.}
            \label{fig:Schematic_of_the_modified_source}
        \end{figure*}
        
    Based on the above derivation of WIE, we develop a simple experiment consisting of three parts: First, we produce polarization entangled photons in the singlet state \eqref{eq:singlet}, see Figure \ref{fig:Schematic_of_the_modified_source} a). Second, we distribute them to two measurement stations where they are detected under different settings, see Figure \ref{fig:Schematic_of_the_modified_source} b). Third, the signals are fed into a coincidence logic (AND-gate) and recorded together with the settings, see Figure \ref{fig:Schematic_of_the_modified_source} c). 
    
    Because of its ubiquity and didactic value in illustrating the necessity of erasing which-way information when generating entanglement, we employ a Sagnac-type spontaneous parametric down-conversion source to produce entangled photons \cite{First_Sagnac_Entanglement,Kim2006,Hentschel2009,Entangled_photon-pair_sources}; however, any commercially available polarization-entangled source may be used.
    We use a $\SI{405}{\nano \meter}$ laser with mode cleaning (PH) and polarization control (PBS and HWP) which is split (dPBS, HWP) and focussed (L2) from both sides into a Type-II phase-matched periodically poled Potassium Titanyl Phosphate (ppKTP) crystal. This creates two opposing photon pair emission processes. For the clockwise directed source, a dHWP helps the pump to satisfy the crystals polarization conditions. Entanglement is created by mixing the outputs of the two processes $\ket{H,V}$ and $\ket{V,H}$ at the same dPBS, rendering them indistinguishable. The output is then described as a superposition.     
        \begin{equation}
          \ket{\Psi} =   \sqrt{w}\ket{HV}+e^{i \xi}\sqrt{1-w}\ket{VH}
            \label{eq:sagnacstate}
        \end{equation}

    For anti-diagonal pump polarization, i.e.,~$\xi = \pi$ and $w = 1/2$, that is set by the HWP, we generate the singlet state \eqref{eq:singlet}.The detection setup consists of motorized polarization filters (mP) implementing projective polarization measurements, followed by single-mode optical fibers that guide the photons to avalanche photodiode single-photon detectors (SD). A coincidence unit identifies joint detection events within a $\SI{2}{\nano \second}$ coincidence window.
    This coincidence window compensates for timing jitter and optical path differences. Singles and coincidence counts are recorded in real time. For testing WIE, only the coincidence counts provided in the data set are required.

\subsection{Data Analysis}
    To determine the Wigner value the joint probabilities used in Equation \eqref{Wigner_value} need to be reconstructed from counting coincidences at different measurement settings. Experimentally, we estimate the joint probabilities using the ratio of coincidences $C$ for a specific measurement setting relative to the total coincidences. For example, $P_{++}^{\hat{a}\hat{b'}}$, the joint probability for both photons of a photon pair to pass through their respective polarization filter with the measurement settings $\hat{a}$ and $\hat{b'}$ is given by 
        \begin{equation}
                \label{pos_prob}
                P_{++}^{\hat{a}\hat{b'}} = \frac{C_{++}^{\hat{a}\hat{b'}}}{C_{++}^{\hat{a}\hat{b'}}+C_{+-}^{\hat{a}\hat{b'}}+C_{-+}^{\hat{a}\hat{b'}}+C_{--}^{\hat{a}\hat{b'}}}.
        \end{equation}
    Here, $C_{++}^{\hat{a}\hat{b'}}$ are the coincidences measured with settings $\hat{a}$ and $\hat{b'}$ for Alice and Bob respectively.
    The index $+$ (plus) denotes a polarization filter is aligned with the measurement direction, transmitting photons polarized along this axis. The index $-$ (minus) corresponds to a $90^\circ$ rotation, transmitting the orthogonal polarization component and blocking the original one. The denominator of Equation \eqref{pos_prob} is the total number of coincidence counts, summed over all polarization outcomes for the measurement settings $\hat{a}$ and $\hat{b}'$.
    Simultaneous measurement of  both $+$ (plus) and $-$ (minus) would be possible with polarizing beam splitters and additional detectors instead of filters.
    
    The remaining joint probabilities required for Equation \eqref{Wigner_value} follow analogously.
    
    \begin{tcolorbox}[colback=gray!5, colframe=black!60, title=Exercise 3]
    The recorded data used in this experiment is available \href{https://www.iqoqi-vienna.at/publications/downloads/wigner-dataset}{here}. The dataset may be analyzed for personal study and educational purposes using software of choice, such as Python.
    \end{tcolorbox}
    
 \section{Angular Dependence Study}
    To fully understand under which choice of basis-angles quantum predictions differ from classical predictions characterized by WIE, several correlation curves were recorded. Acquiring one correlation curve required approximately 24 hours, and the measurements were therefore automated. In principle, the experiment could also be repeated by manually changing the measurement settings. 

    \subsection{Dependence on the Relative Angle}
    \label{sec:M1}
        For the first of three measurements, we keep the absolute angle of both measurement-bases fixed, i.e.,~$\theta = \ \theta_{Alice} =\theta_{Bob} = \ket{V}$ and vary the relative angle $\phi$, thereby increasing the angular spread of the measurement directions. 
        
            \begin{figure*}[!t]
                \centering
                \includegraphics[width=\textwidth]{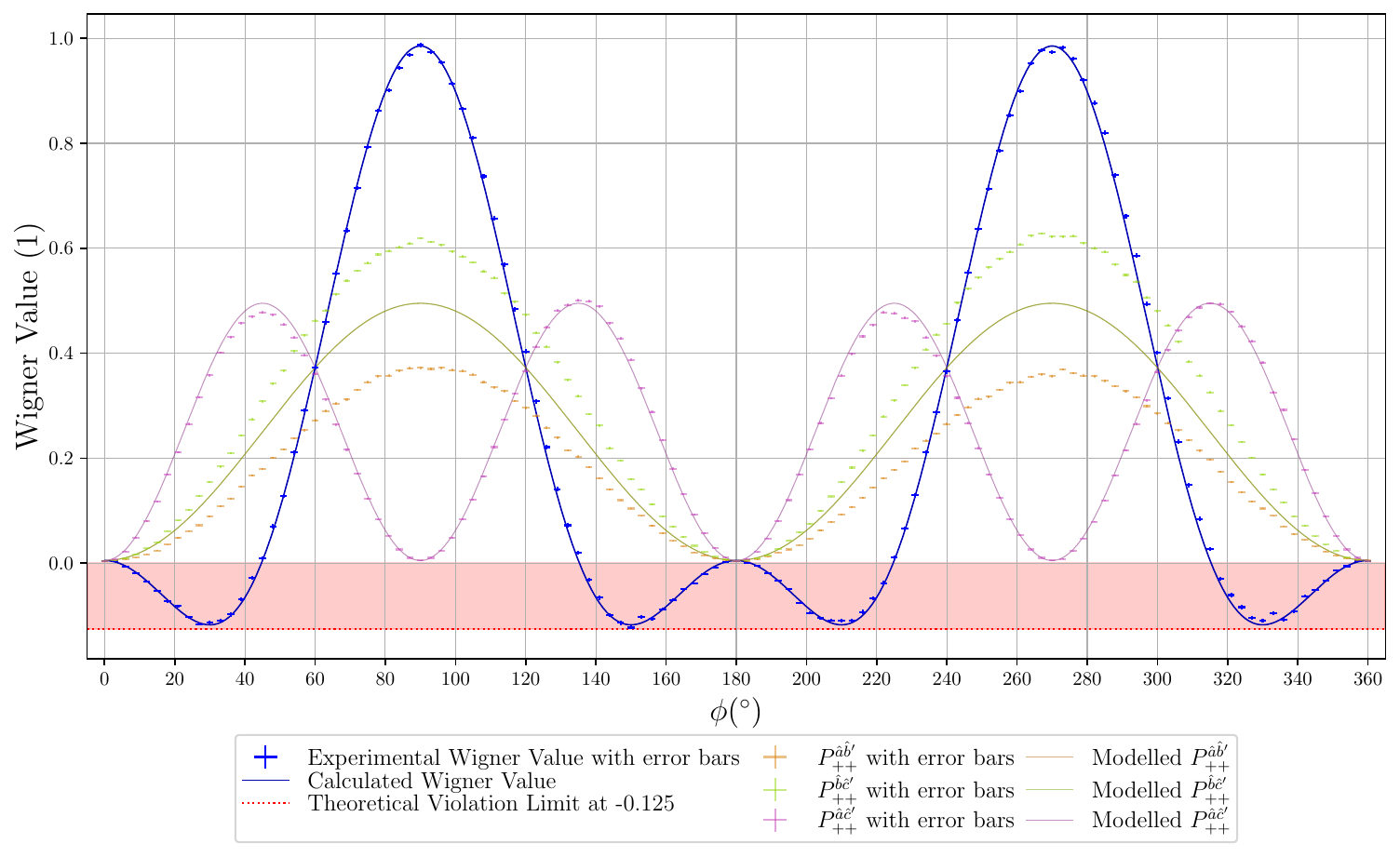}
                \caption{Plot of Wigner Value $\mathcal{W}$ versus symmetric relative angles. The absolute angles are fixed $\theta_{Alice} =\theta_{Bob}$. A Wigner value of $-0.123$ with a significance of $34 \, \sigma$ was observed. The model state, Equation \eqref{eq:modelstate}, successfully fits the Wigner value curve, although the individual probability curves deviate from the measured data points.}
                \label{fig:Plot of Wigner Value to symmetric relative angle}
            \end{figure*}
            
        In Figure \ref{fig:Plot of Wigner Value to symmetric relative angle}, we plot the experimental Wigner Value $\mathcal{W}$ and its component individual probabilities over a broad range of relative angles $\phi$. We then fit these curves by calculating $\mathcal{W}$ for the empirical model state
        
        \begin{equation}
             \rho = v\ket{\Psi}\bra{\Psi} + \frac{1-v}{4}\mathbbm{1},
             \label{eq:modelstate}
        \end{equation}

        where the weight $w$ appearing in Equation \eqref{eq:sagnacstate} and the visibility $v$ of the isotropic noise model are used as fit parameters, while fixing $\xi = \pi$.
        The model provides a minimal quantum-mechanical description of the experiment, but does not include a full detector-level simulation.Consequently, it reproduces the Wigner value curves well but cannot fully account for the partial probability curves. The remaining deviations can be attributed to polarization-dependent losses and other experimental asymmetries.
        Note that varying $w$ physically corresponds to changing the ratio of photon pairs propagating in the clockwise and counterclockwise directions that enter the fiber couplers. The visibility $v$ accounts for white noise in both couplers. A least-square fit for this measurement yields $w = 0.50$ and $v = 0.98$.

         \begin{tcolorbox}[colback=gray!5, colframe=black!60, title=Fidelity]
            The fidelity quantifies how ``close" a state is to the desired target state. For a state $\rho$ and a pure target state $\ket{\Psi}$, this is equivalent to their overlap. Hence, the fidelity becomes
            \begin{equation}
                \label{Fidelity}
                \mathcal{F} = \operatorname{Tr}\left(\rho\ket{\Psi}\bra{\Psi}\right)
            \end{equation}
            Notice that $0\leq \mathcal{F} \leq 1$, where $F = 0$ implies orthogonal states and $F = 1$ holds for identical states.

        \end{tcolorbox}
        
        The fidelity \cite{NielsenChuang2010} between the model state $\rho$ and the singlet Bell state, Equation \eqref{eq:singlet}, is $\mathcal{F} = 0.985$.
        This high fidelity indicates strong agreement between the experimentally reconstructed state and the ideal singlet state, and can be interpreted as a measure of the source quality. We show four segments of $45^\circ$, for a full period, where negative values for $\mathcal{W}$ occur.
        
    \subsection{Dependence on Symmetric Absolute Angles}
    \label{sec:M2}
        Subsequently, the dependence of WIE on two symmetric absolute angles of the measurement basis is studied ($\theta_A = \theta_B$). In practice, this relates to fully turning both polarization filters synchronous, to increase the absolute angles. The relative angle of the measurement bases $\phi$ is kept constant at $30^\circ$ for maximal angular spacing.
        
            \begin{figure*}[t]
                \centering
                \includegraphics[width=\textwidth]{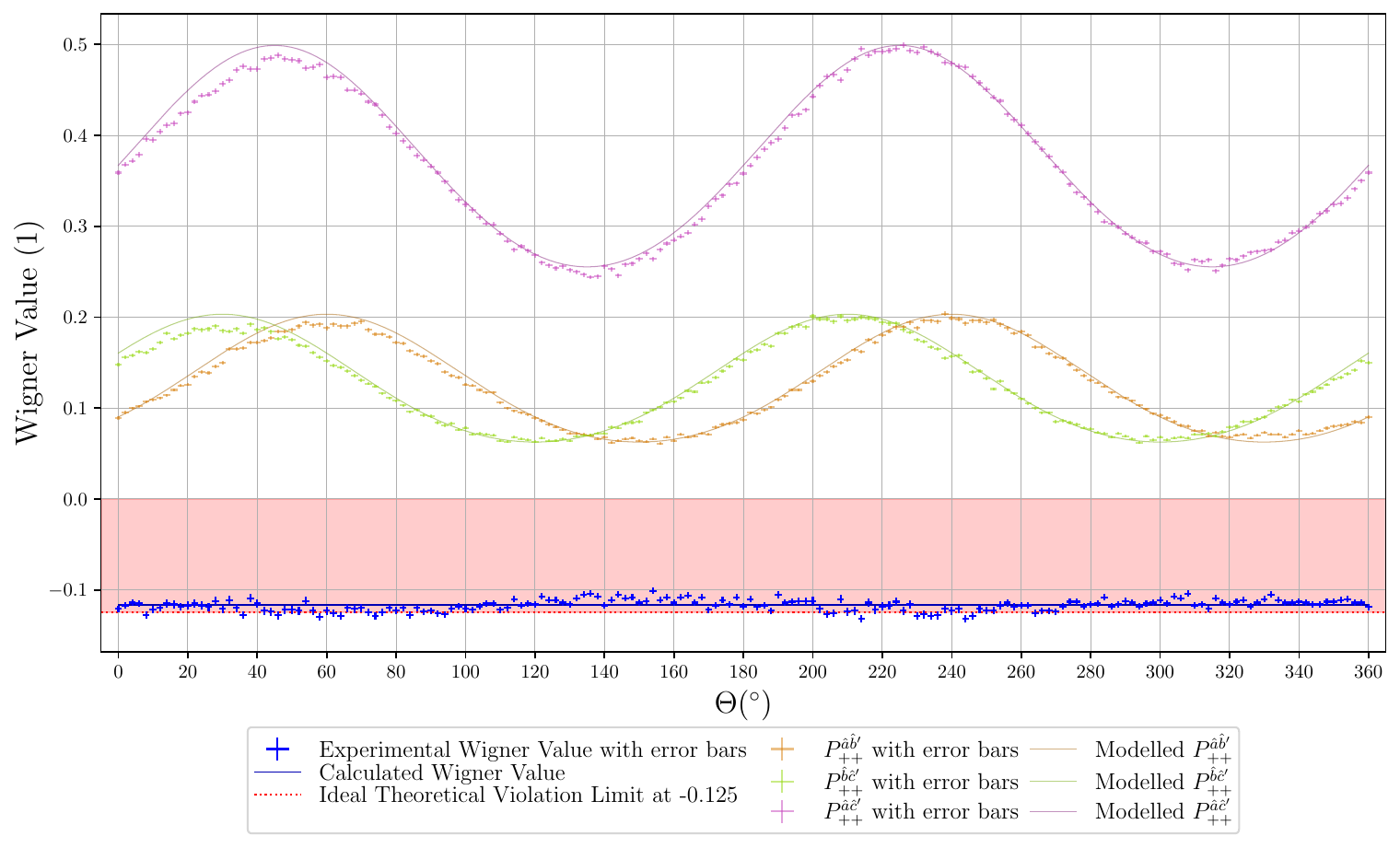}
                \caption{Plot of Wigner Value $\mathcal{W}$ to symmetric absolute angles ($\theta_A = \theta_B$). The relative angles $\phi$ of Alice and Bob are fixed at $30^\circ$. The mean measured violation of Wigner’s inequality is $-0.117$, corresponding to a statistical significance of $30 \, \sigma$, in close agreement with the theoretical maximum violation for this measurement configuration of $-0.125$.}
                \label{fig:Plot of Wigner Value to symmetric absolute angles}
            \end{figure*}
            
        After fitting this data, we obtain $w = 0.36$ and $v = 0.99$, which yields a fidelity of $\mathcal{F} = 0.978$. The results are presented in Figure~\ref{fig:Plot of Wigner Value to symmetric absolute angles}. Since the prepared state is not a Bell singlet state, all curves exhibit sinusoidal deviations from the ideal behavior.

    \subsection{Dependence on Asymmetric Absolute Angles}
    \label{sec:M3}
        Finally, we scan WIE for asymmetric absolute angles by fixing one basis at a relative angle of $\phi = 30^\circ$ and rotating the other basis. Once, the measurement basis of Alice is held in place ($\theta_{Alice} = 0^\circ$), whilst the basis of Bob is rotated (varying $\theta_{Bob}$), and vice versa. This measurement procedure is therefore conducted twice.

            \begin{figure*}[!t]
                \centering
            
                \includegraphics[width=\textwidth]{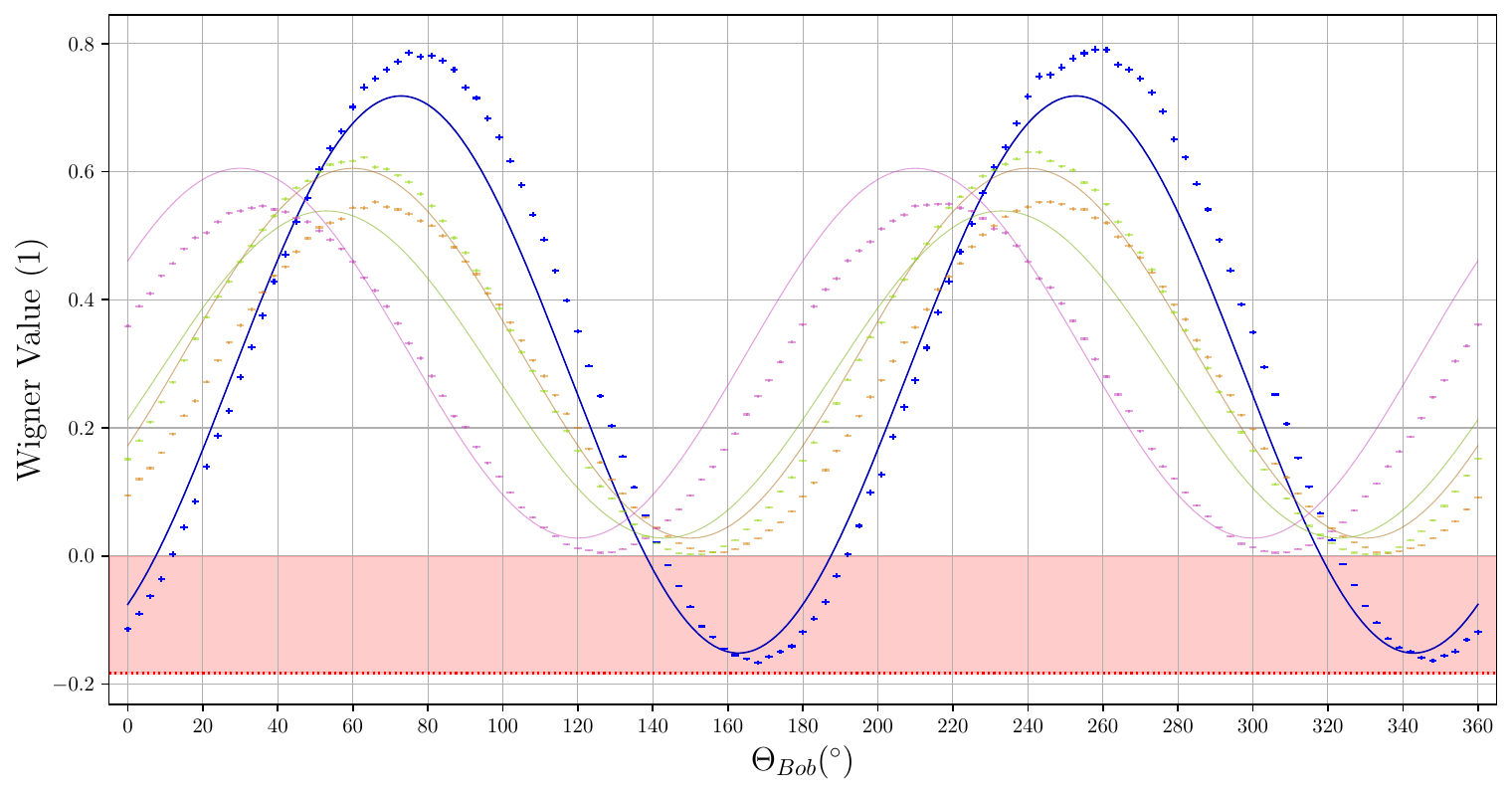}
            
                \includegraphics[width=\textwidth]{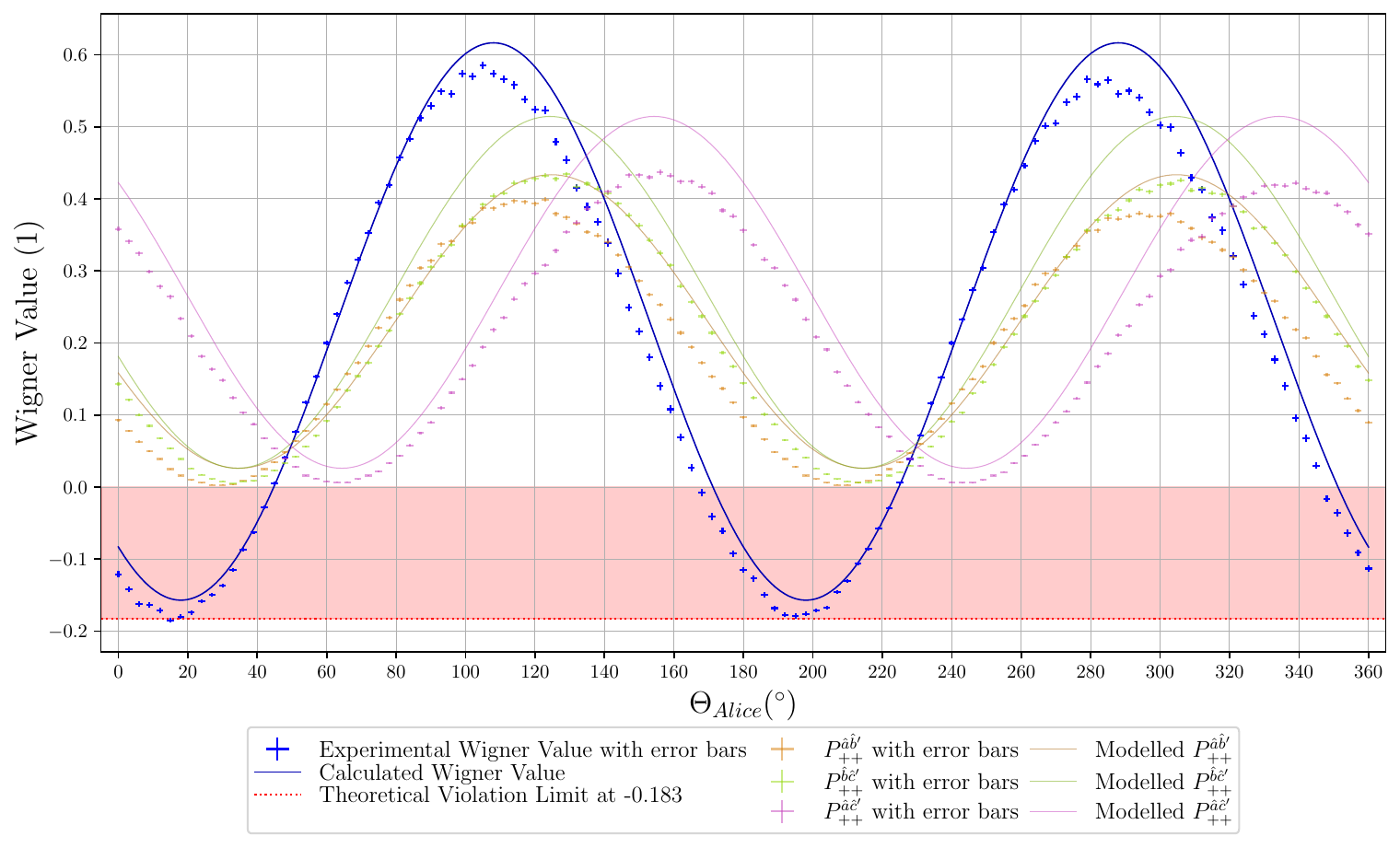}
            
                \caption{Top: Plot of Wigner value $\mathcal{W}$ versus asymmetric absolute angle with Alice fixed. A minimum value of $\mathcal{W} = -0.167$, corresponding to a statistical significance of $56 \,\sigma$, was obtained.\\ Bottom: Corresponding plot with Bob fixed. A minimum value of $\mathcal{W} = -0.185$, corresponding to a statistical significance of $54 \,\sigma$, was obtained. In both cases, the empirical model does not fully reproduce the maxima of the measured curves.}
                \label{fig:wigner_fixed_comparison}
            \end{figure*}
        The least-square fit in Figure \ref{fig:wigner_fixed_comparison} (top) yields $w = 0.35$ and $v = 0.89$, corresponding to a fidelity of $\mathcal{F} = 0.896$.

        For the fit in Figure \ref{fig:wigner_fixed_comparison} (bottom) a slightly different state and density operator are used, where $w = 0.41$ and $v = 0.90$ give a fidelity of $\mathcal{F} = 0.914$.

\subsection{Results}
   
        In the first measurement, Section \ref{sec:M1}, the strongest violations of WIE occur for $\phi = \pm 30^\circ$, and for a phase shift of $180^\circ$. As outlined in Section \ref{sec:Expected Violations of Wigner's inequality}, this phenomenon appears when the measurement settings are chosen to achieve maximal symmetric angular separation within the given configuration. This highlights the sensitivity of quantum correlations to basis alignment. A  Wigner value of $-0.123$ was observed at the second minimum in Figure \ref{fig:Plot of Wigner Value to symmetric relative angle}, corresponding to a $34 \, \sigma$ deviation. Statistical uncertainties and corresponding significance levels are computed assuming Poisson-distributed photon counts, with Gaussian error propagation applied to the derived Wigner parameter \eqref{Wigner_value}.
        
        The second measurement, Section \ref{sec:M2}, demonstrates that violations of Wigner’s Inequality are invariant under unitary rotations, underscoring the robustness of quantum non-locality under certain transformations .As seen in Figure \ref{fig:Plot of Wigner Value to symmetric absolute angles}, WIE is violated for every measurement point. Compared to the constant ideal violation limit, with unit fidelity, of -0.125, sinusoidal variations of the Wigner Value arise. The mean Wigner value is $-0.117$ with a statistical significance of $30 \, \sigma$. The observed deviations in $\mathcal{W}$ arise from the sinusoidal behavior of the individual probabilities. Small phase shifts in the experimental data prevent perfect cancellation, resulting in a residual sinusoidal modulation.
        
        The third measurement, presented in Section \ref{sec:M3}, involves asymmetric rotations and shows a gradual reduction in violation strength until quantum and classical correlations become indistinguishable.
        Minimum values of $\mathcal{W} = -0.167$ and $\mathcal{W} = -0.185$ are obtained, in the top and bottom plots of Figure \ref{fig:wigner_fixed_comparison}, respectively, corresponding to deviations from the classical bound of $56 \,\sigma$ and $54\,\sigma$. In comparison to the second measurement, not all measurement points taken violate WIE when testing with asymmetric absolute angles.
        
        For a full rotation of one polarizer, negative Wigner values are observed only within two angular regions of approximately $55^\circ$, separated by half a period. Outside this range, the data do not allow a conclusive statement regarding the non-local behavior of the system.
        \pagebreak
        \clearpage
        
        This illustrates the fragility of the system under improper measurement conditions.  
        One could expect the top and bottom plots of Figure \ref{fig:wigner_fixed_comparison} to be identical, as the measurements conducted were mirror-symmetrical. 
       
        The remaining variations can be attributed to two factors. The phase shift of $30^\circ$ in the two figures is attributed to the order of the measurement directions, representing a geometric occurrence that is induced by the structure of WIE, Equation \eqref{eq:Wigner_inequality_equ_sin}. Moreover, the different amplitudes, fitted parameters and fidelities are due to the slight natural drift in the setup, which influences the ratio between $\ket{HV}$ and $\ket{VH}$ coupled in by the optical fibers.

\section{Conclusion}

    In this work, we present an accessible derivation of Wigner’s Inequality and implement a widely-used experimental platform to test the compatibility of measured correlations with classical and quantum models. Three measurement configurations were analyzed, yielding correlation curves with distinctly different behaviors depending on the chosen settings.
    
    The results were evaluated in terms of violations of Wigner’s Inequality and their statistical significances, enabling a direct comparison with the classical bound. We observe angular regions with statistically significant violations, with statistical significances between $30\,\sigma$ and $56\,\sigma$, while other measurement settings remain consistent with both quantum and local realistic models within experimental uncertainty. This highlights the strong dependence of the observed violation on the measurement basis and the importance of systematic parameter scans.
    
    Beyond the physical results, the main contribution of this work is pedagogical: it provides a complete framework linking theoretical derivation, experimental implementation, and open data analysis. This makes the experiment well suited for undergraduate teaching, allowing students to reproduce the analysis and directly observe violations of a Bell-type inequality in measured data.
    
    Overall, Wigner’s Inequality serves as a transparent and intuitive platform for demonstrating quantum correlations and for introducing students to authentic methods of experimental quantum physics and data analysis.

\section*{Acknowledgments}
M. R. and D. S. thank Paul Erker for providing valuable comments and suggestions that improved the clarity of the paper. M. R. is also grateful to Eugenia Benech Charbonnier and Robert Kindler for their practical advice and helpful discussions during the initial phases of the laboratory work. This work was supported by the Open-Access-Fund of the Austrian Academy of Sciences, and by the University of Vienna via the QUESS project (Quantum Experiments at Space Scale).
For the purpose of open access, the author has applied a CC-BY public copyright licence to any Author Accepted Manuscript version arising from this submission.

 \bibliography{citations.bib}

\end{document}